\documentclass{jaa}
\newcommand{\A}{\emph{AstroSat}}

\usepackage{natbib}
\usepackage{amsmath}
\usepackage{mathtools}
\usepackage{graphicx}
\usepackage{caption}
\usepackage{subcaption}
\usepackage{aas_macros}
\usepackage{hyperref}
\usepackage{float}

\begin{document}\sloppy

%%paper title
%%For line breaks \\ can be used within title 
%\title{Improvements to event selection in data analysis pipeline for CZTI Imager of AstroSat}
\title{A generalized event selection algorithm for AstroSat CZT Imager data}

%%author names are separated by comma (,) 
%%use \and before the last author name 
%%use a * along with the number separated by comma
%% for the  author for correspondence
%%\textsuperscript{number} is used for affiliation
%%\affilOne, \affilTwo etc., upto \affilTwentyfive is possible
%%Please note the first letter after \affil is capitalised in the command
%%

\author{A. Ratheesh\textsuperscript{1,2,3}, 
A. R. Rao\textsuperscript{1,4},    
N.P.S. Mithun\textsuperscript{5}, 
S.V. Vadawale\textsuperscript{5}, 
A. Vibhute\textsuperscript{4}, 
D. Bhattacharya\textsuperscript{4}, P. Pradeep\textsuperscript{6}, S. Sreekumar\textsuperscript{6} and 
V. Bhalerao\textsuperscript{7}
}
%\author{AUTHOR1\textsuperscript{1}, AUTHOR2\textsuperscript{1} and AUTHOR3\textsuperscript{2,*}}
\affilOne{\textsuperscript{1}Tata Institute of Fundamental Research,
            Homi Bhabha Road, Colaba, Mumbai, 400005, India.\\}     
\affilTwo{\textsuperscript{2} Department of Physics, Tor Vergata University of Rome, Via della Ricerca Scientifica 1, I-00133 Rome, Italy\\}
\affilThree{\textsuperscript{3} INAF - IAPS, Via Fosso del Cavaliere 100, I-00133 Rome, Italy.\\}
\affilFour{\textsuperscript{4}Inter University Centre for Astronomy \& Astrophysics, Post Bag 4, Ganeshkhind, Pune,  411007, India.\\}
\affilFive{\textsuperscript{5}Physical Research Laboratory, Navrangpura, Ahmedabad, 380009, India.\\}
\affilSix{\textsuperscript{6}Vikram Sarabhai Space Centre, Kochuveli, Thiruvananthapuram,  695022, India.\\}
\affilSeven{\textsuperscript{7}Indian Institute of Technology Bombay, Mumbai, India\\}

%%escape two column mode for title, affiliation and abstract
%%by giving \twocolumn command as shown

\twocolumn[{

\maketitle

%%include \corres to print the corresponding author Email id
\corres{ajay.ratheesh@roma2.infn.it, ajayratheesh@gmail.com}

%%include \msinfo for
%%manuscript information such as
%%received, revised and accepted dates
%%
\msinfo{}{}

%%abstract
\begin{abstract}
%The Cadmium Zinc Telluride (CZT) Imager on-board AstroSat is a hard X-ray imaging spectrometer operating in the energy range of 20 – 100 keV. It also acts as an open hard X-ray monitor above 100 keV capable of detecting transient events like the Gamma-ray Bursts (GRBs). Additionally, the instrument has sensitivity to measure hard X-ray polarisation in the energy rage of 100 – 400 keV for bright on-axis sources like Crab and Cygnus X-1 and bright GRBs. The current CZTI pipeline caters mainly to the requirements of the spectral and timing measurements of the on-axis sources and for the additional sciences like transient search and polarisation studies, customised tweaking of the pipeline and the use of custom built software are necessary. Further, handling of the cosmic-ray induced noise need to be done subjectively based on the science requirements. Here we present a generalised pipeline developed to cater to multiple use of CZTI data.The efficacy of this pipeline is reviewed by examining the Poissonian behaviour of the events and the signal to noise ratio of detecting GRBs.
The Cadmium Zinc Telluride (CZT) Imager on board AstroSat is a hard X-ray imaging spectrometer operating in the energy range of 20 $-$ 100 keV. It also acts as an open hard X-ray monitor above 100 keV capable of detecting transient events like the Gamma-ray Bursts (GRBs). Additionally, the instrument has the sensitivity to measure hard X-ray polarization in the energy range of 100 $-$ 400 keV for bright on-axis sources like Crab and Cygnus X-1 and bright GRBs. As hard X-ray instruments like CZTI are sensitive to cosmic rays in addition to  X-rays, it is required to identify and remove particle induced or other noise events and select events for scientific analysis of the data. The present CZTI data analysis pipeline includes algorithms for such event selection, but they have certain limitations. They were primarily designed for the analysis of data from persistent X-ray sources where the source flux is much less than the background and thus are not best suited for sources like GRBs. Here, we re-examine the characteristics of noise events in CZTI and present a generalized event selection method that caters to the analysis of data for all types of sources. The efficacy of the new method is reviewed by examining the Poissonian behavior of the selected events and the signal to noise ratio for GRBs.
\end{abstract}


}]
%%close the twocolumn escape here

%%include \doinum{number}for the DOI number in the header
%%include \volnum{number} for the volume number in the header
%%include \year{yyyy} for  year of publication in the header
%%include \pgrange{num--num} page range of article in the header
%%include \artcitid{num} for the article citation id
%%include \lp to print last page of the article
%%include \setcounter{page}{pagenum} for the exact starting page of the article

\doinum{12.3456/s78910-011-012-3}
\artcitid{\#\#\#\#}
\volnum{000}
\year{0000}
\pgrange{1--}
\setcounter{page}{1}
\lp{1}

\section{Introduction}
CZT Imager (hereafter CZTI) is one of the four co-aligned instruments used for pointed observations in the Indian multi-wavelength astronomical satellite \A\, \citep{Singh_et_al.-2016-SPIE}.  CZTI is a hard X-ray instrument operating above 20 keV \citep{Bhalerao_et_al.-2017-JApA}.  It uses passive collimation and coded mask imaging to make spectral and timing measurements in the 20 -- 100 keV region. Above these energies the collimators and shield become increasingly transparent and  the instrument can be used as an all sky monitor to detect transient events like Gamma-ray Bursts (GRBs) and measure their spectral, timing and, most importantly,  polarisation properties \citep{Rao_et_al.-2016-ApJ, Chattopadhyay_et_al.-2017-arXiv}.   Though the basic CZT detectors are sensitive to only upto 200 keV, the Compton scattered events can be used to measure the total energy upto 380 keV. Further, some individual detector elements, called pixels, happened to be of low gain and hence can be used to extend the energy range of the instrument to 700 keV (see Chattopadhyay et al. and Abhay et al., this volume).
%\cite{Chattopadhyay_submeV,Abhay_submeV}). 

Non-focussing hard X-ray instruments are generally  background dominated and in CZTI the coded mask technique effectively measures the background and the source
simultaneously within its primary field of view. At higher energies, where CZTI has some attractive additional scientific features like the spectro-polarimetric study of GRBs, the background can be highly variable and unpredictable, limiting the sensitivity of measurements. There are, however, several design features in CZTI, though used earlier in several hard X-ray instruments but rarely simultaneously in any instrument, which can  be very useful in reducing and eliminating background. \A\, is in a low inclination (6$^\circ$) low Earth orbit, thus  making the satellite only skim the surface of the high background South Atlantic Anomaly (SAA) region in most of the satellite orbits. This drastically reduces the proton induced radioactivity. Among the currently operating hard X-ray instruments only the NuSTAR satellite is in such an equatorial orbit. Secondly, CZTI uses a limited amount of shielding because it is used for imaging only up to 100 keV. This results in a much lower amount of high-Z materials thus reducing the effective induced background. The most important feature of CZTI, however, is the continuous availability of the time-tagged information for each of the registered ionising event: energy of the event, its location in the detector plane, and the time of arrival (correct to 20 $\mu$s). Since hard X-ray detectors are single photon counting devices, the effect of background can be  understood and perhaps eliminated by examining the data in multiple dimensions (spatial, temporal and energy) and demanding a strict adherence to the Poisson nature of random events. Thus a  pixelated detector with information of time, energy and position of interaction can distinguish between particles, particle induced noises, electronic pixel noises and genuine X-rays. This is due to the fact that the interaction process of the photons are different from particles, and their statistical properties in space, time and energy differ.

A pixelated detector like CZTI is an ideal detector to search for short transients ($<$1 s) like short Gamma Ray Bursts (sGRBs), counterparts to gravitational wave events and fast radio bursts (FRBs) \citep{Rao_msdomain_2018JApA...39....2R}. The standard CZTI pipeline primarily provide the analysis of data for on-axis sources and the search for such transient events is currently done by fine-tuning the pipeline parameters. CZTI reports the detection of around 80 GRBs per year. However, most of these detections are based on trigger times obtained from other instruments like Gamma Ray Burst monitor onboard Fermi Gamma-ray Space Telescope (Fermi-GBM), Burst alert monitor onboard Neil Gehrels Swift Observatory (Swift-BAT) and so on. These detections are also prone to manual fine tuning of some parameters to get the maximum signal to noise ratio for that GRB. Some efforts have been made to search for transients in CZTI data, however such untriggered searches for transients require a substantial reduction of noise, without which such efforts will result in a large fraction of false alarms. Other than the electronic noises, the main source of noise in a semi conductor detector in space is cosmic particle induced noises. Particles get scattered within the detector crystal resulting in a large amount of ionisation. Apart from triggering pixels at time of arrival of a particle, pixels become `noisy' for a further time period due to large deposition of charge. This time period is of the order of few tens of milli seconds, which affects the search for transients especially at short time scales. 

In this work, we examine the properties of `events' in CZTI to segregate noise events from science analysis to to develop a source independent algorithm for reducing the cosmic ray induced noises and to improve the performance of the instrument. Primarily we study the characteristics of the cosmic particle induced noises in addition to a better handling of the electronic noises. In a companion paper %\cite{Paul_dphstructures}
(Paul et al. this volume) we presented the results of the study of the data for spatial clustering and found signatures for particle track interactions that can mimic short transients at time scales of a few hundred milli-seconds, as seen in the INTEGRAL PICsIT instrument \citep{Segreto_et_al.-2003-A&A--INTEGRAL_cosmic_rays}. Here we take a holistic view and examine the data in all the available dimensions (spatial, temporal, and energetic) and develop an improved algorithm to segregate the noise from genuine X-ray `events'. The details of the event selection methods employed in the current CZTI pipeline and its inadequacies are discussed in the next section. Temporal, spatial, and spectral characteristics of noise events in CZTI are presented in Section 3 and the details of a new algorithm for removal of these events is provided in Section 4. In Section 5, the efficacy of this algorithm is demonstrated and in the last section, we conclude discussing its advantages.

\section{An overview of CZTI data and event selection in the data analysis pipeline}
\label{sec:CZTIDetails}

CZTI consists of $4$ independent quadrants each having an array of 16 pixellated CZT detector modules ~\citep{Bhalerao_et_al.-2017-JApA}. Each CZT detector module  is of size $\sim 40~\rm{mm} \times 40~\rm{mm}$ and composed of 256 pixels with 2.5 mm pitch. All quadrants of CZTI also include veto detectors, made of CsI(TI), placed underneath the CZT detector plane, which helps in identifying and removing coincidentally detected particle events and high energy gamma rays.    

CZTI operates in different modes, depending on the storage space available, charge particle background and various other external and internal factors. The normal mode (M0) is the default mode of operation in which it records the time tagged information for all the events that are registered. For each event, the recorded data contains the information regarding the detected energy of incident photons (PHA), pixel number, module identification (id), time stamp correct to 20 $\mu \rm{s}$, a flag indicating if the event is from on-board calibration source and the energy deposited in the veto detector if it is a veto-tagged event. This list of events constitute the basic data from CZTI for all subsequent analysis.

Apart from genuine X-ray photons from the source and the X-ray background, charged particles  also contribute to the `events' recorded by X-ray detectors like CZTI. Particles interacting with the detector lose energy continuously and produce tracks in the detector plane. Charged particles interacting with the instrument or spacecraft body can produce secondary particles and X-ray photons, which can also deposit energy in the detectors. For the scientific use of data from instruments, such particle induced events are to be identified and removed by the data processing chain. 

As CZTI is composed of pixellated detectors, each pixel acts as an independent X-ray or particle detector. One charged particle, however, can produce events in many pixels of CZTI at the same time by the interaction of the same particle as well as its secondaries. Due to the finite time required for polling and readout of event details from each pixel, it is possible that these events are recorded with different time stamps, although they occurred at nearly the same time. Hence, such particle interactions will be recorded as successive events having time stamps that differ by zero or by the  time resolution element of CZTI: 20 $\mu s$ (The $ 20 {\rm \,\mu s} $ time bin is  digitally generated;  the on board   electronics is capable of recording events at a  time scale of $\sim$ $ 1 {\rm \,\mu s}$, and  the  time required for polling and readout of an event is a few $\mu$s, much shorter than the CZTI time resolution element). We call these train of events as `bunches', as they are bunched temporally. As the detectors in CZTI are designed for very high count rate applications, they have very short ($\sim 1 ~\mu s$) charge readout time. Hence, when a high energy particle deposits significant amount of charge in a pixel, it can trigger multiple events that are recorded individually. Thus, bunches also include multiple events from one pixel.
It may be noted that as two events that are recorded within 20 $\mu s$ can be Compton scattered events that carry information about the polarization of the source ~\citep{Vadawale_2018NatAs...2...50V,Chattopadhyay_et_al.-2017-arXiv}, bunches are defined as three or more events clustered in time.

Since such bunched series of events are not from the source under observation, an on-board algorithm identifies them and removes them from the event records written in the on-board storage for transmission to ground. For each bunch of events discarded this way, a summary including the number of events in the bunch, total duration, full information for three events (first, second, and the last), detector numbers for additional four events, are recorded instead. For the rest of the events, complete information as mentioned earlier are transmitted to the ground and all further selection of events are carried out by the data analysis pipeline on ground. In most cases the pruned data was found to be an order of magnitude lower in volume  than the noisy data, thus providing a huge advantage in the data transmission time. 

Analysis of in-flight data has shown that often there are residual additional events in time scales of the order of a milli-second after the end of bunched events. These are understood as due to electronic noise induced by the `bunch'. The low energy threshold of the instrument is kept just above the electronic noise ($\sim$ 20 keV) to facilitate the collection of a large number of Compton scattered events. This has the effect of making the pixels prone to some false triggering. Some pixels become quite noisy and they are suppressed by ground commands.  Further, some pixels become noisy after the cosmic ray charge deposition and hence some lingering noise is also seen after the bunches induced by cosmic rays. 

In the CZTI level-1 to level-2 data processing chain, `cztbunchclean' task takes care of this post-bunch event cleaning. Bunches are defined as series of events which have time stamps of successive events differing by 20 $\mu s$ (the time resolution of CZTI) or less. Events that are part of bunches are removed from the event list by  `cztbunchclean'. Further, events for certain duration after each bunch are also removed by this task. As it is observed that bunches with less number of events are generally localized in the detector plane, post bunch events removal is done differently for short bunches and long bunches. For bunches with length less than 20, all events within  1 ms (\textit{T2}) duration after the bunch, registered in the same detector where the bunch occurred are discarded. For bunches with length more than 20, all events within 1 ms (\textit{T3}) are discarded irrespective of the detector. The algorithm also includes a provision to ignore events after the bunch irrespective of the bunch length for certain duration (\textit{T1}); however, this is not used in the default processing and is not recommended. It may be noted that there is provision to vary the threshold bunch length and time scales from the default values.

Although all the events removed in this manner are expected to be particle induced events, a very small fraction of genuine X-ray events coinciding with the bunch duration or post-bunch duration also get removed. Hence, the live time of the instrument needs to be corrected for this. The `cztbunchclean' task also computes the live time losses incurred by the removal of particle shower duration, which is used in subsequent stages to correct the exposure times. The reduction in exposure is of the order of $\sim 1-2 \%$. It is found that approximately 90\% of the recorded events are due to events discarded as bunches and they cause this 1-2\% reduction in exposure. The typical background rate, per quadrant, is 150 to 200 s$^{-1}$.

Apart from these particle induced train of events, other spurious events occur due to certain pixels having higher noise. During the ground testing and in-flight commissioning phases of the instrument, pixels that have extremely high noise have been disabled. Similar exercise is carried out from time to time during the mission operation. However, during each observation, it is likely that some additional pixels misbehave and generate spurious events, but not high enough to decide to disable that pixel. In such cases, events from these `hot pixels' also need to be removed from further analysis. It is to be noted that  near room temperature pixelated semiconductor detectors like CZT are prone to increased noise in some pixels and these noisy pixels are isolated and removed by ground software \citep[see, for example,][for a description of the noise reduction method for Swift/BAT.]{segreto_2010A&A...510A..47S}.\\

In CZTI data analysis software, the task named `cztpixclean' identifies and discards events from these noisy pixels. This is carried out in two steps. In the first step, pixel-wise histograms for each quadrant for the entire observation is computed. For a total of 4096 pixels per quadrant, and with an average count rate of $\sim150$ $counts/s$, none of the pixels are expected to deviate beyond $5$ $\sigma$ from the mean due to the Poisson statistics that a photon counting instrument should follow. Hence, pixels that show a deviation more than $5$ $\sigma$ from the total average in a Detector Plane Histogram (DPH) are considered as "noisy" pixels and are removed from any further analysis. At least a data set with at least $2000$ $s$ are required to get enough statistics per pixel to allow such an analysis. This process is done iteratively and the iteration is continued until all remaining pixels have counts within the five sigma limit. In this process, the number of the events removed as noisy pixel events depends on the number of noisy pixels and counts per noisy pixel in that observation. In general around one percent of the pixels are flagged as "noisy pixels".
%In the next step, flickering detectors and pixels that have noisy events only within certain duration of the observation are identified and events from them are discarded only for that duration. Such flickering pixels and detectors are identified as those having events \textbf{count rate} higher than a certain limit for each time interval considering the maximum deviations expected from a Poisson distribution with the mean count rates. The algorithm removes data from pixels for time bins of \textit{pix\_tbinsize} where they register more than \textit{pix\_count\_thresh} events and removes data from detectors for time bins \textit{det\_tbinsize}  having more than \textit{det\_count\_thresh} events. The default configuration (hereafter \textit{cztpipeline\_lowthresh\_run}) ignores one second bins of pixels having more than 2 counts and detectors having more than 25 counts. These thresholds are determined by assuming the average count rates observed by CZTI which is dominated by background in case of observations of persistent X-ray sources. 

In the next step, flickering detectors and pixels that have noisy events only within certain duration of the observation are identified and events from them are discarded only for that duration. Such flickering pixels and detectors are identified as those having count rates higher than a certain limit, considering the maximum deviations expected from a Poisson distribution with the mean count rates. As the average count rates observed by CZTI is dominated by background, it is expected to remain constant in case of observations of persistent X-ray sources and thus the default threshold rates (hereafter \textit{cztpipeline\_lowthresh\_run}) for pixels and detectors required for the algorithm are estimated assuming the typical background rates. Thus, the algorithm discards events from pixels for one second time bins that register more than 2 counts/s and from detectors that register higher than 25 counts/s.

However, this assumption is not valid in the case of transients like GRBs, where the count rates increase significantly above the background rates. In such cases, this part of the algorithm in `cztpixclean' with the default parameters will tend to remove significant fraction of source events in addition to the noise events. Thus, for purposes like search for transients and analysis of GRBs with CZTI observations, this part of the event selection in the data analysis pipeline is bypassed by providing high values for the detector and pixel count thresholds (hereafter \textit{cztpipeline\_highthresh\_run}). While this ensures that no source events are rejected, some of the spurious events remain in the final list of selected events. Various values based on the flux level of the GRBs are manually set by trial and error to get the maximum signal to noise ratio and to get a stable background. Although it is possible to optimize the threshold parameters based on the count rate of an already identified GRB, it will not be possible to do that in case of requirements like blind search for transients.

The event selection algorithms employed in CZTI data analysis pipeline as described here is found to be effective to identify and remove most of the noise events for observations of persistent X-ray sources. With the low threshold configuration in `cztpixclean', it is estimated that the contribution from the residual noise events in the clean event files is less than 10\% of the statistical error due to the background (see Section 5). For the analysis of bright transients like GRBs, however, the software parameters had to be tweaked depending on the brightness of the GRBs so that the source counts are not suppressed as noise.
%, but this is not applicable for the analysis of bright transients like GRBs.
Considering these limitations in the event selection algorithm in the current data analysis pipeline, here we re-examine the characteristics of the noise events in CZTI and propose improvements to these methods that overcome these limitations.

\section{Noise characteristics at millisecond time scales}

\begin{figure*}[ht!]
\centering
\includegraphics[width=\hsize]{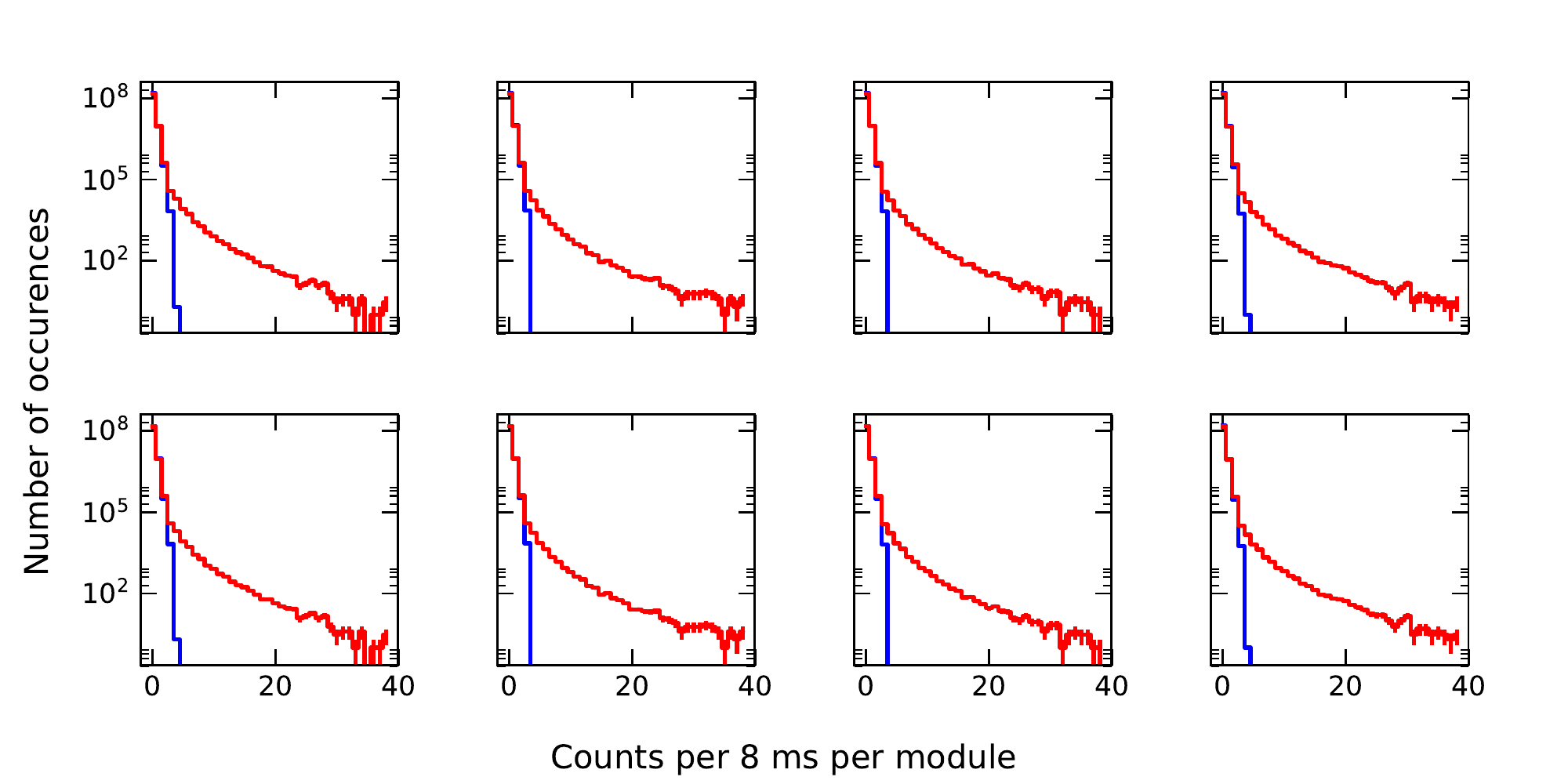}
\caption{Histogram of counts per 8 ms per detector module for obsID1 (top) and obsID2 (bottom). Columns from the left shows quadrant 0,1,2 and 3, respectively. The blue line shows the expected Poissonian distribution  and the red line shows the observed count distribution.}
\label{msplot_beforeclean}
\end{figure*}

We have investigated the events at milli-second time scales to study the characteristics of the noise events. %identify the source of non-Poissonian behaviour. 
The main motivation to explore milli-second time scales is the observation that the lingering effects of cosmic particles  can extend upto a few tens of milli-seconds, even though the primary interactions and energy deposition are in nano-seconds. To perform this study, we use the output event file obtained from `cztbunchclean' with no post bunch cleaning (\textit{T1}$=$0, \textit{T2}$=$0,and \textit{T3}$=$0). Though the new algorithm is tested on several data, for a systematic study that is presented here we have selected two data sets for a detailed analysis: AstroSat observation IDs 9000000618 (hereafter obsID1) and 9000000276 (hereafter obsID2). The data sets are of   sufficiently long duration   so that they cover  the observed diurnal variation in the background. We further flag the `noisy' pixels from this analysis, as their source is already identified as electronic noises. Since the cosmic rays interact locally at the module level, the noise triggered by it should also be locally clustered at the place of interaction of the particle. Hence we probe the time scale of lingering noise at the module level. We generated lightcurves at 0.5 ms, 1.0 ms, 2.0 ms, 4.0 ms,  8.0 ms and 16.0 ms for each module. From the mean count-rate we estimate the expected  counts from a Poisson distribution  and compare it with the observed counts. The estimated Poissonian counts and observed counts histogram for each module is then added to get statistics at the quadrant level. Since the background of CZTI shows a slow orbital variation, we divide the data into chunks of 100 s intervals. The observed and expected counts from all these chunks are summed up to get the final histogram. \\

\begin{figure}[hb!]
\centering
\includegraphics[width=\hsize]{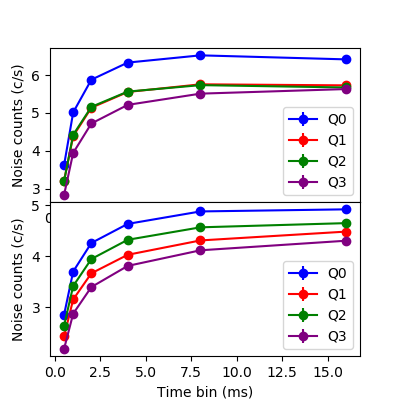}
\caption{The noise counts at different time scales for all the quadrants. The noise counts are estimated from the deviation from Poissonian. Top and bottom plots are for obsID1 and obsID2. Different quadrants are labelled within the plot.}
\label{noisecounts_timescale}
\end{figure}

An example of such a histogram for 8.0 ms binning is shown in Fig. \ref{msplot_beforeclean}. A consistent behaviour is seen for   both the obsIDs as well as for  all quadrants of CZTI. It can be seen that there are a large number of occasions when counts in excess of 10 are found, while the expected number from Poisson distribution is practically zero.  We note here that any deviation in the Poisson distribution by slow variation in the counts due to orbital background variation is explicitly taken care of by calculating the expected Poisson distribution for the each 100 s chunk of data. Further, by taking the counts for each module we essentially reduce the average expected rate and thus enhancing the contrast due to noise because the noise variations are expected to be localised to each module. 
We can use these histograms to estimate the residual noise counts by integrating the observed histogram and subtracting the expected Poisson counts. Thus for the 8.0 ms binned data, the estimated noise count rates in all the quadrants are 23.53 c/s and 17.9 c/s in obsID1 and ObsID2, respectively, for the 8.0 ms binned data. We use this technique to get an  estimate of the clustered noise events at various time scales. The estimated noise event rates at 0.5 ms, 1.0 ms, 2.0 ms, 4.0 ms, 8.0 ms and 16.0 ms timescales are  plotted in Fig. \ref{noisecounts_timescale}. Again a consistent trend is seen in the two obsIds and across quadrants: though there is about 20\% variation among the different sets of data. It demonstrates that the noise events have a time scale of several ms and most of the noise events are captured when we examine the data at 8 ms time scale.
%(see Fig. \ref{noisecounts_timescale}) and 
Hence we chose 8 ms for further estimating the amount of noise events. 
%Fig. \ref{msplot_beforeclean} shows the deviation from the Poissonian at 8 ms time scale for obsID1 and obsID2. The estimated noise count rate is 23.53 c/s and 17.9 c/s in obsID1 and ObsID2. 
%Since the particles also follow a Poissonian statistics, we assume that the noise generated by it should also follow Poissonian statistics. We fit the distribution with two Poissonian functions for genuine and noise events. 

\begin{figure}[ht!]
\centering
\includegraphics[width=\hsize]{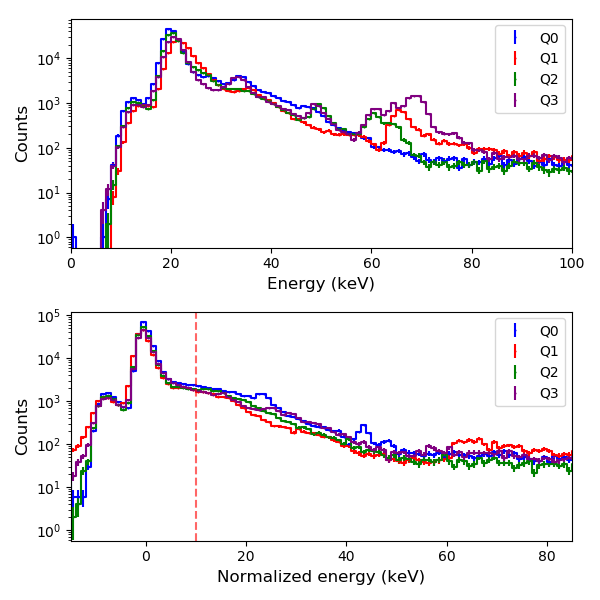}
\caption{Top: The spectral energy distribution of events selected from 8 ms bins of module wise data when the  observed counts go beyond 10 counts per bin (deemed to be noise events, see text). Multiple peaks are seen in the distribution. Bottom: The same distribution when the energy scale is re-normalised for each module by taking the start point as the LLD of that module.  A clear peak is seen near the LLD and the red vertical line shows the energy threshold used for further selection criteria. Different quadrants are indicated by the labels within the plots. The  data corresponds to obsID2.}
\label{msnoise_energydist}
\end{figure}

\begin{figure}[ht!]
\centering
\includegraphics[width=\hsize]{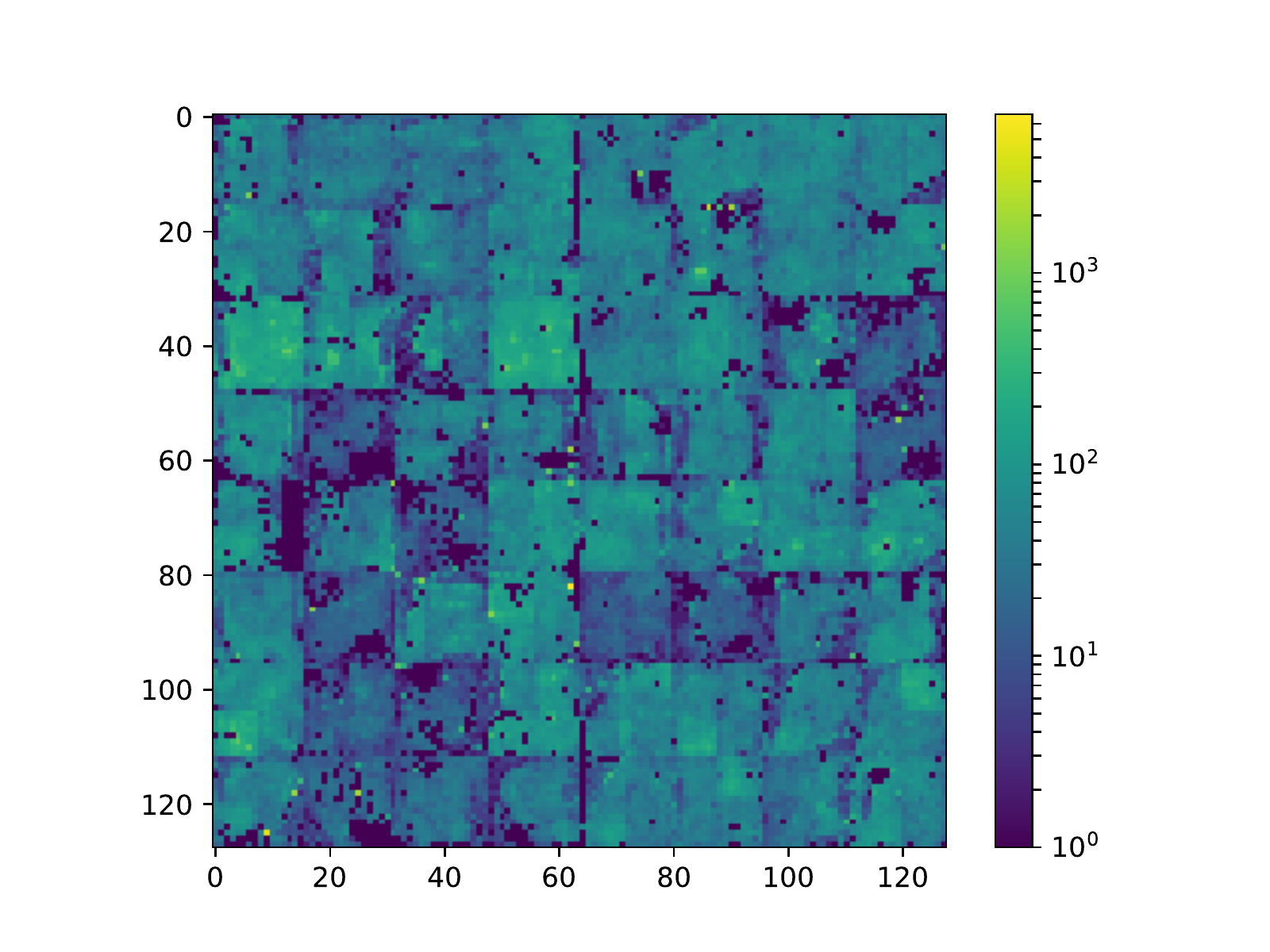}
\caption{The detector plane histogram (DPH) of the noise events selected according to the criterion that the integrated counts over 8 ms bins} goes beyond 10 counts. The plotted data correspond to obsID2.
\label{msnoise_dph}
\end{figure}

\begin{figure*}[ht!]
\centering
\includegraphics[width=\hsize]{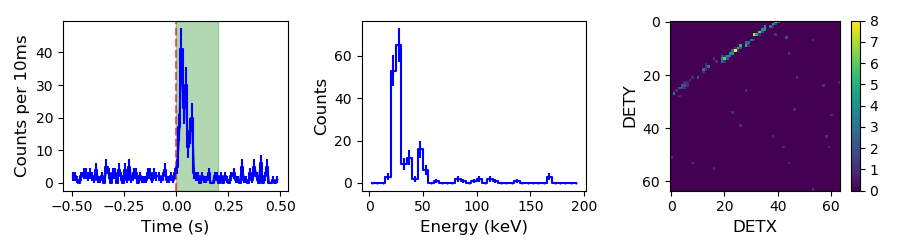}
\caption{Left: The light curve binned at 0.01 s showing the post bunch noise after a bunch. The time of the bunch is indicated as red vertical line. The green shaded region shows time selection for the energy distribution and the DPH. middle and right: The energy distribution and DPH for the post bunch noise.}
\label{bunch_track}
\end{figure*}

Since the expected Poissonian count histogram falls to zero beyond 10 counts per 8 ms per module (see Fig. \ref{msplot_beforeclean}), all the counts detected beyond it have to be  noise counts. We used these counts to study their distribution in energy across the detector plane to further investigate their characteristics. When we made a plot of the  energy spectra of  these events, it  showed multiple peaks at different  energies  (top panel of Fig. \ref{msnoise_energydist}). On a closer inspection of the data we realised that these peaks are module dependent and hence we scaled the energy scale of the plot to the Lower Level Discriminator (LLD) value of each module (we note here that in CZTI there are no pixel wise LLD, rather there is only a module wise LLD: each module having only one analog to digital converter). Different modules have their LLDs in the range 15 keV to 65 keV.
%where different modules have their LLD We further normalized the energies of each of these events by subtracting the LLD of the pixel at which the event was recorded.
In the distribution of the normalized energy (taking the energy scale from the LLD of each module), we see that there is only a single peak near the LLD, thus signifying that the noise events are clustered around the electronic LLD of the modules. 
%rather than the calibrated energy scale.  the multiple peaks disappeared and beyond 10 keV, only the background events dominate. This indicates that the noise events are clustered 
Hence we keep a threshold of 10 keV from the LLD of each module to identify the clustered noise events if they satisfy further conditions as outlined in the next section. 10 keV is an adequate value to cover the entire range of post bunch noises and the genuine events from being flagged based on trial and error on different data sets.

We also examined the detector plane histogram (DPH) of the above mentioned noise events. The counts are uniform across the quadrant except for a few remaining ’noisy’ or ’flickering’ pixels as seen in the detector plane histogram (DPH) of these events (Fig. \ref{msnoise_dph}). This indicates that these are electronic noises, however they are uniform and hence not associated to just a few ’noisy’ pixels. We conclude that these extra noise events are caused by the lingering effects of cosmic ray induced bunches in all active pixels and they appear uniform in the DPH. 
%They occur predominantly near the LLD of the module and Since the bunches are large in number (∼ 70 per second) they appear uniform in the DPH.

Now we examine the lingering effects after a particle interaction in the detector plane. When a particle interacts and triggers multiple pixels within 20 $\mu$s, then they are registered as a bunch as mentioned above. We find excess counts for few tens of milliseconds just after the interaction of many `bunches'. However these are mainly found for `bunches' with more than 15 events. Fig. \ref{bunch_track} shows the post bunch noise associated with the bunch, the lightcurve showing excess counts after the bunch, and energy distribution which shows excess counts near the LLD. The time interval for the energy distribution and the DPH are indicated as green shaded part in the light curve. The time of the `bunch' is indicated by the vertical red line. From the DPH the post bunch noise along the track left by the particle can be seen clearly. In most of the cases the time scale of post bunch effects is around a few tens of milli-seconds, however in some extreme cases, the time scale can go as large as 250 ms. The plotted data corresponds to quadrant 0 of obsID2. Similar properties in energy and temporal regimes for post bunch noises in comparison to the non-Poissonian noise identified above indicate that the post bunch noises are the main source of the  non-Poissonian noise.

\section{A new event selection algorithm}
\begin{figure*}[hbt!]
\centering
\includegraphics[width=\hsize]{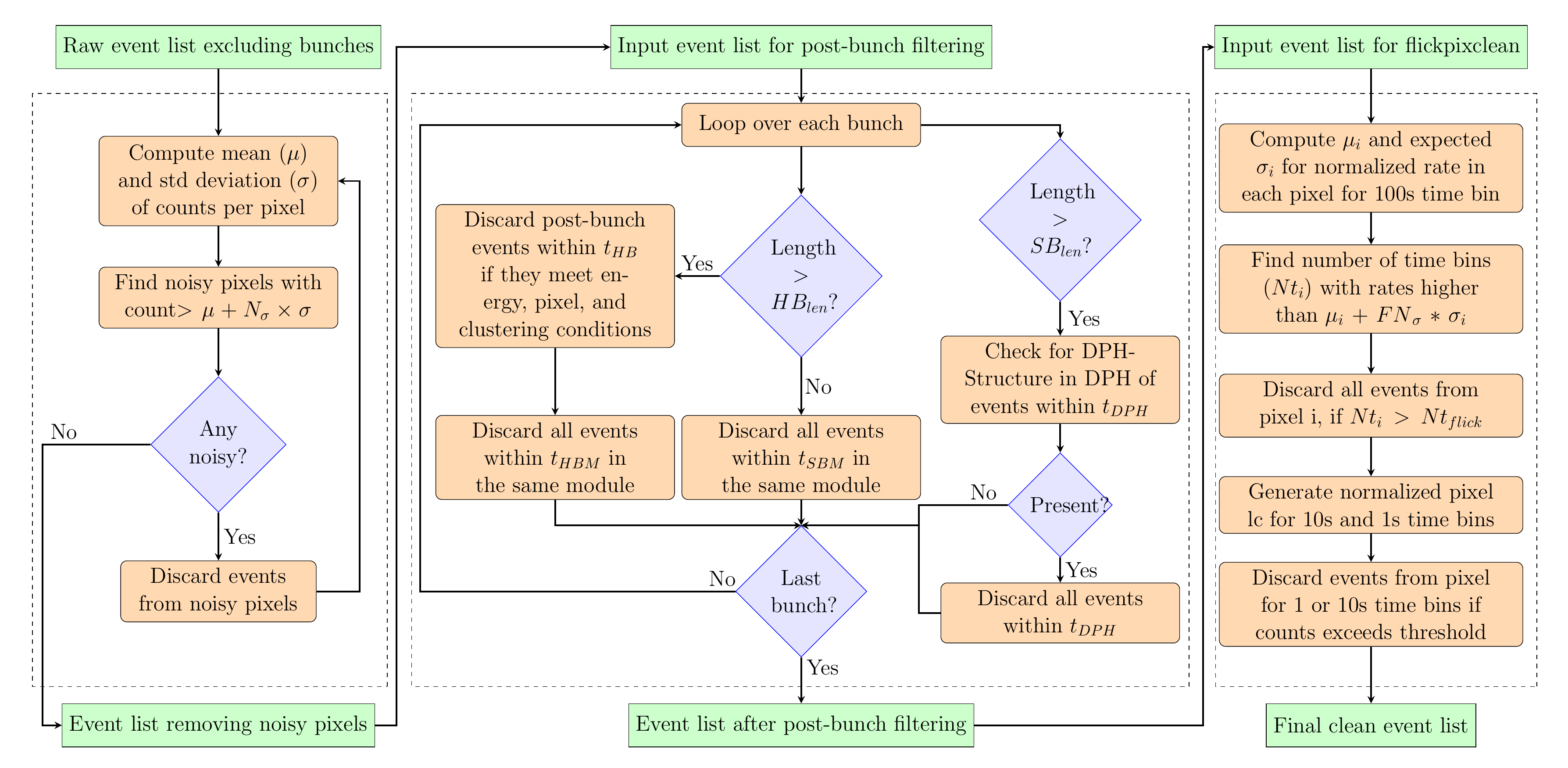}
\caption{Flow chart outlining the new event selection algorithm sequentially. The abbrevations used are given in the bracket for different parameters in Table \ref{table_parameters}.}
\label{nc_flowchart}
\end{figure*}

Based on all the above findings, we have formulated a  strategy for removing the noise events. This essentially replaces the routines `cztbunchclean' and `cztpixclean'  in the present analysis pipeline. The algorithm is divided into four different subsequent tasks.
%The gross 'noisy' pixels are flagged at the beginning since the post bunch noise are uniform across the detector, and the noisy pixels can be identified as outliers in the DPH. Several improvements, however, are made as compared to the existing algorithm like treating the low gain pixels and edge pixels separately (see below).  The second and third steps involve removal of post bunch noise based on two improved algorithms in place of that in \textbf{'cztbunchclean'}. Finally at the end we flag 'flickering' pixels at different time scales. Search for 'flickering' pixels are done after the removal of cosmic particle noises, as the time scales of both are comparable, and the chances of genuine pixels with a lingering cosmic particle effect getting flagged as a "flickering" pixel is high.

At first we remove the gross noisy pixels from the detector plane histogram (DPH) of a quadrant, as it is the predominant component of noise in the detector. This is done in an iterative manner. All the pixels in a quadrant which register counts above $5$ sigma from the mean are classified as `noisy' pixels and are removed from the analysis. Since normal pixels and `low gain' pixels have a difference in the count rate, the flagging is done separately for them. Since the edge pixels in a module have less geometric area, all the counts in DPH are normalised by the geometric area of the particular pixel. The mean and sigma are recalculated after the first iteration to catch and exclude fainter `noisy' pixels. The iteration is repeated until no further pixels are caught as `noisy'. The deviation from the mean (5 $\sigma$) is parametrised as \textit{noisypixsigmathresh}. This step is similar to the algorithm employed in `cztpixclean' of the CZTI pipeline, but with some refinements such as  handling the low gain pixels separately. This same method of flagging pixels is additionally used in the final event file for identifying noises in the `neighbouring double' events used for polarimetry. Two events occurring within in the time resolution of the instrument are termed as `double' events, and such events in neighbouring pixels are termed as `neighbouring double' events. Since the corner pixels angles register less events, we flag the outliers separately for the corner and side double events. We further split the flagging process for low gain pixels and normal pixels as the count rate observed in them are different. If a pixel is found noisy for any of the segregated DPHs, then no further Compton events are used from that pixel for polarimetry.

In the second stage of event selection, we identify and remove post-bunch noise events. This task is carried out in two steps based on the number of events in the bunch.
%\textbf{In the second task (or main step), we identify and remove the post-bunch noises associated to bunches. This task is made on two further steps}. 
As bunches with more than 100 events span more than one module, the module identification becomes tricky as they are identified from $7$ module numbers among all the bunch events. When such a bunch occurs it may trigger pixels across at least 2 modules and hence can disturb a large number of pixels. However, these bunches can be identified from the total number of events in a bunch. At present we keep a threshold of 100 events (parametrised as \textit{superbunchsize}) to distinguish such bunches. These bunches are classified as `super bunches'. After identifying these bunches we perform the \textit{DPHclean} method outlined in Paul et al. (this volume) to check if there is a \textit{DPHstructure} or spatial clustering in the DPH for the next $100$ ms (parametrised as \textit{DPHtime}) from the start time of the bunch. If a clustering is found, that time interval is excluded for that quadrant. The fractional live-time during this time are updated accordingly. In the next part of the post bunch clean, we include bunches with events greater than 15 events (parametrised as \textit{heavybunchsize}) as well. These bunches are called as `heavy bunches', and the rest of the smaller bunches are termed as `small bunches'. These particle tracks that trigger less number of bunch events, can be locally identified in a module or two. For such a bunch, 4 further criteria determine if these post bunch events need to be excluded. These events are excluded if all the four criteria are met at the same time

\begin{itemize}
    \item the event is within $250$ ms (parametrised as \textit{heavybunchtime}) after the bunch,
    \item the energy of the event is less than LLD + 10 keV. The LLD here is the LLD of the pixel in which the event is found,
    \item the event is in the same module as the bunch, or in the neighbouring 4 columns and 4 rows of pixels adjacent to that module,
    \item the event is clustered in $8$ ms with another event that also follows the above three criteria.
\end{itemize}
%If an event follows all the above  four conditions, then that event is excluded from further analysis.
Furthermore, all events from the same module or neighbouring pixels are excluded for $5$ ms (parametrised as \textit{heavybunchtimedet}) or $1$ ms (parametrised as \textit{smallbunchtimedet}) for `heavy bunches' or `small bunches'. The pixel exposure values of each pixels are corrected accordingly.

The final task of the algorithm is to identify and remove events from flickering pixels. Since pixels can flicker from very short time scales to long time scales, we apply two strategies to find and exclude events from flickering pixels. The first strategy is to find pixels that show non-Poissonian behaviour with respect to time and thus identified from the deviations in the individual pixel light curves. The light curves are binned at $100$ seconds for each pixels and the deviations from the mean are checked to find the flickering pixels.   A bin size of $100$ s is appropriate to find the flickering pixels as it gives enough statistics per bin. The significance of deviation (parametrised as \textit{flickersigma}) and the number of allowed times for the deviation (parametrised as \textit{flickernumtimes}) are decided by the total exposure of the observation. Since the overall background of CZTI follows a trend, the pixel wise light curves are normalized by the total light curve for the quadrant. This will essentially avoid flagging short time increase in count-rate due to genuine GRBs detection, as flickering episodes because such increase will be more or less uniformly distributed across the detector plane.
Further, the count rate in each bin is corrected for the fractional exposures calculated from the good time intervals (GTIs) of each quadrants. Pixels caught as flickering will be excluded for the entire observation period. The second strategy to find flickering pixels is looking for non-Poissonian behaviour in the DPH at shorter time scales, i.e. 1 s and 10 s. Depending upon the mean count rate in that time interval, the pixel counts which are not probable to occur at least once are calculated, starting from 2 counts, and all the pixels that register counts above that are removed for that particular interval. Pixel exposures are updated accordingly after this step. Table. \ref{table_parameters} summarize all the parameters and the respective default values used in the new algorithm. The flow chart given in Fig \ref{nc_flowchart}, outlines the entire algorithm. The abbreviations used in the flow chart for visual representation are given in the Table. \ref{table_parameters}.

\begin{table}[h]
\caption{Parameters of the event selection algorithm. \textit{flickersigma} and \textit{flickernumtimes} are based on the exposure of an observation. The range of values used are given in the table.}
\label{table_parameters}
\centering   
\renewcommand{\arraystretch}{1.5}
\begin{tabular}{c c} 
\hline\hline  
Parameter & Default value \\
\hline
\textit{noisypixsigmathresh ($N_{\sigma}$)} & 5 \\
\textit{superbunchsize ($SB_{len}$)} & 100 \\
\textit{DPHtime ($t_{DPH}$)} & 100 ms  \\
\textit{heavybunchsize ($HB_{len}$)}  & 15 \\
\textit{heavybunchtime ($t_{HB}$)}  & 250 ms \\
\textit{heavybunchtimedet ($t_{HBM}$)} & 5 ms \\
\textit{smallbunchtimedet ($t_{SBM}$)} & 1 ms \\
\textit{flickersigma ($FN_{\sigma}$)} &  5-8   \\
\textit{flickernumtimes ($Nt_{flick}$)} & 1-3 \\
\hline
\hline
\end{tabular}
\end{table}
%\subsection{Optimization of parameters in noise clean algorithm}
\section{Results}
We now employ the algorithm described in the previous section to obtain clean events for CZTI observations. The efficacy of the new event selection strategy is evaluated by examining any residual clustering in temporal, spatial, and spectral domains for the cleaned events. Finally, we use data from detected gamma ray bursts to quantify the improvement in signal to noise ratio with this method.  

\subsection{Clustering at millisecond time scales}
\begin{figure*}[ht!]
\centering
\includegraphics[width=\hsize]{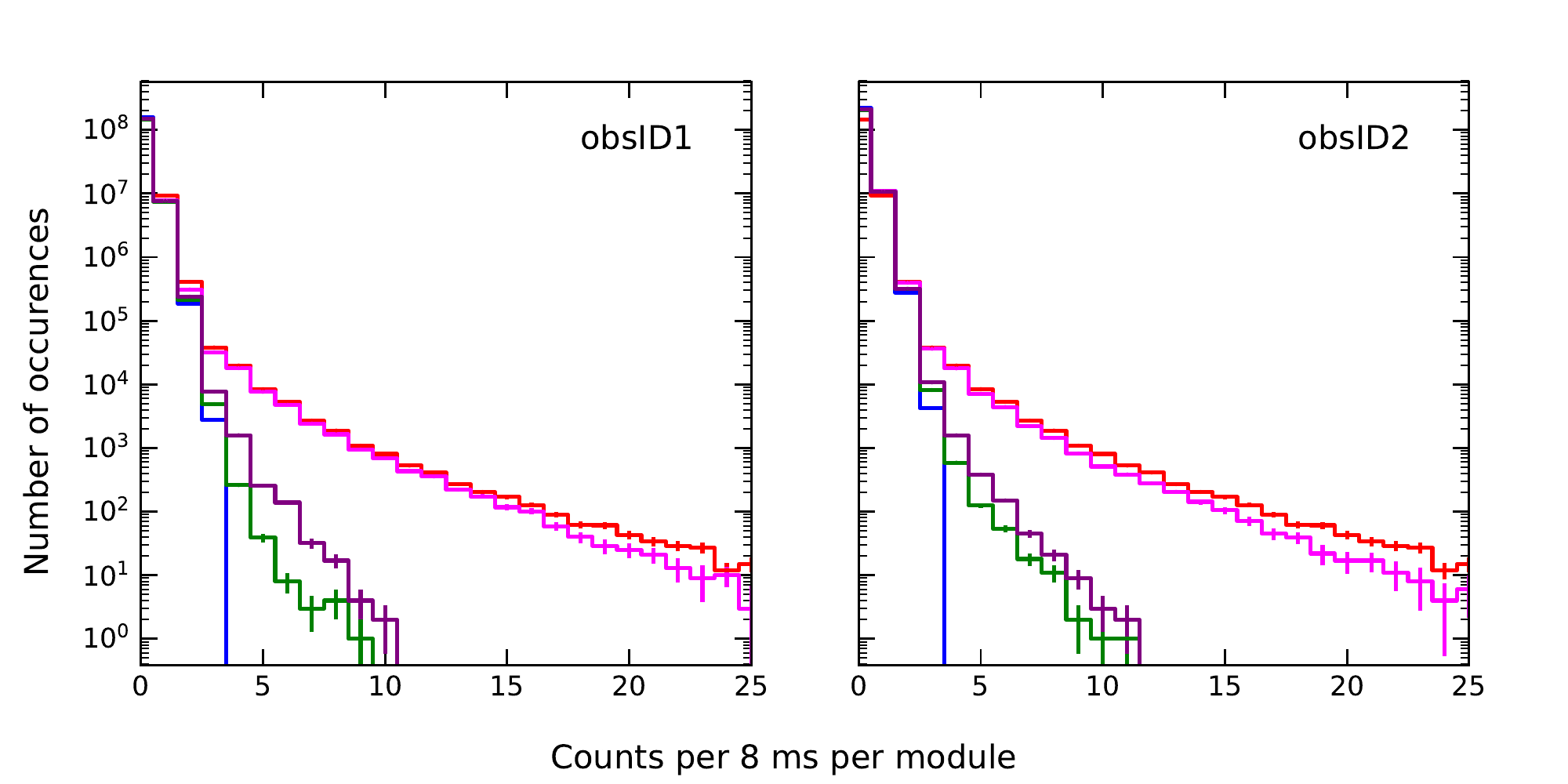}
\caption{Histogram of counts per 8 ms per detector module for obsID1 (left) and obsID2 (right) after the new event selection algorithm (green), as well as \textit{low\_thresh\_config} (purple) and \textit{cztpipeline\_highthresh\_run} (magenta) products.
The plotted data corresponds to quadrant 0.}
The blue line shows the expected Poissonian counts and the red line shows the observed counts without any post bunch clean.
\label{msplot_afterclean}
\end{figure*}

We re-examine the non-Poissonian behaviour at 8 ms time scale in the cleaned event files to quantify the reduction in temporally clustered noise. We find that the noise events calculated from the difference in the observed and the expected count rates decreased by an order of magnitude. The noise count rate after the cleaning is 3.33 c/s and 3.72 c/s in obsID1 and ObsID2 in comparison to 23.53 c/s and 17.9 c/s before the cleaning. We also compare these noise count rates with the results of the both the cztpipeline run configurations (as outlined in section 2) and for both the observations. For the \textit{cztpipeline\_lowthresh\_run} the noise count rate is 5.16 c/s and 4.39 c/s in obsID1 and ObsID2, and in \textit{cztpipeline\_highthresh\_run} is 20.43 c/s and 14.96 c/s in obsID1 and ObsID2. Fig. \ref{msplot_afterclean} shows the expected Poissonian counts in quadrant 0, the observed counts before and after noise clean, and also for the \textit{cztpipeline\_lowthresh\_run} and \textit{cztpipeline\_highthresh\_run} of the cztipipeline. The other three quadrants also show similar behaviour as in Fig. \ref{msplot_afterclean}. From Fig. \ref{msplot_afterclean} and from the noise count rates, it is evident that the current algorithm decreases the number of noise events better than the \textit{cztpipeline\_lowthresh\_run} of the cztpipeline, and at least by a factor of 5 with respect to the \textit{cztpipeline\_highthresh\_run}.

\subsection{Spatial clustering and \textit{DphStructures}}
Using the \textit{DPHclean} algorithm discussed in \cite{Paul_dphstructures} (this issue)
%\cite{Paul_dphstructures}, 
we examine the level of spacial clustering within a bin size of 100 ms at different steps event selection algorithm. This \textit{DPHclean} algorithm is previously used within the event selection algorithm to exclude time intervals after a bunch of size 100 events (see section 4), but here we use it on the entire data set to quantify any spatial clustering in the remaining selected events. A bin size of 100 ms was used since this time scale is identical to the time scale of post bunch noises. DPH for every 100 ms starting from the start time to the stop time is generated and checked for spatial clustering using the above mentioned algorithm. Those DPHs where spatial clustering was found are termed as \textit{DphStructures}. In this section we try to quantify the rate of \textit{DphStructures} in event files at different steps of the noise clean algorithm. Since post bunch noises are the prominent form of spatially clustered noises, the rate of \textit{DphStructures} indicates the amount of residual post bunch noise remaining in the data. We find that the the rate of \textit{DphStructures} after the `noisy' pixel clean is 0.93 per second and 0.95 per second in obsID1 and obsID2. After the \textit{postbunchclean1} it reduced to 0.42 per second and 0.48 per second for obsID1 and obsID2. Further after \textit{postbunchclean2} it reduces to 0.22 per second and 0.19 per second for obsID1 and obsID2 respectively. After \textit{flickpixclean} it again reduced to 0.08 per second and 0.07 per second for obsID1 and obsID2. This indicates that the noise due to \textit{DphStructures} reduces by an order of magnitude after the noise clean algorithm.\\
\subsection{Spectra of clean events}
\begin{figure}[hb!]
\centering
\includegraphics[width=\hsize]{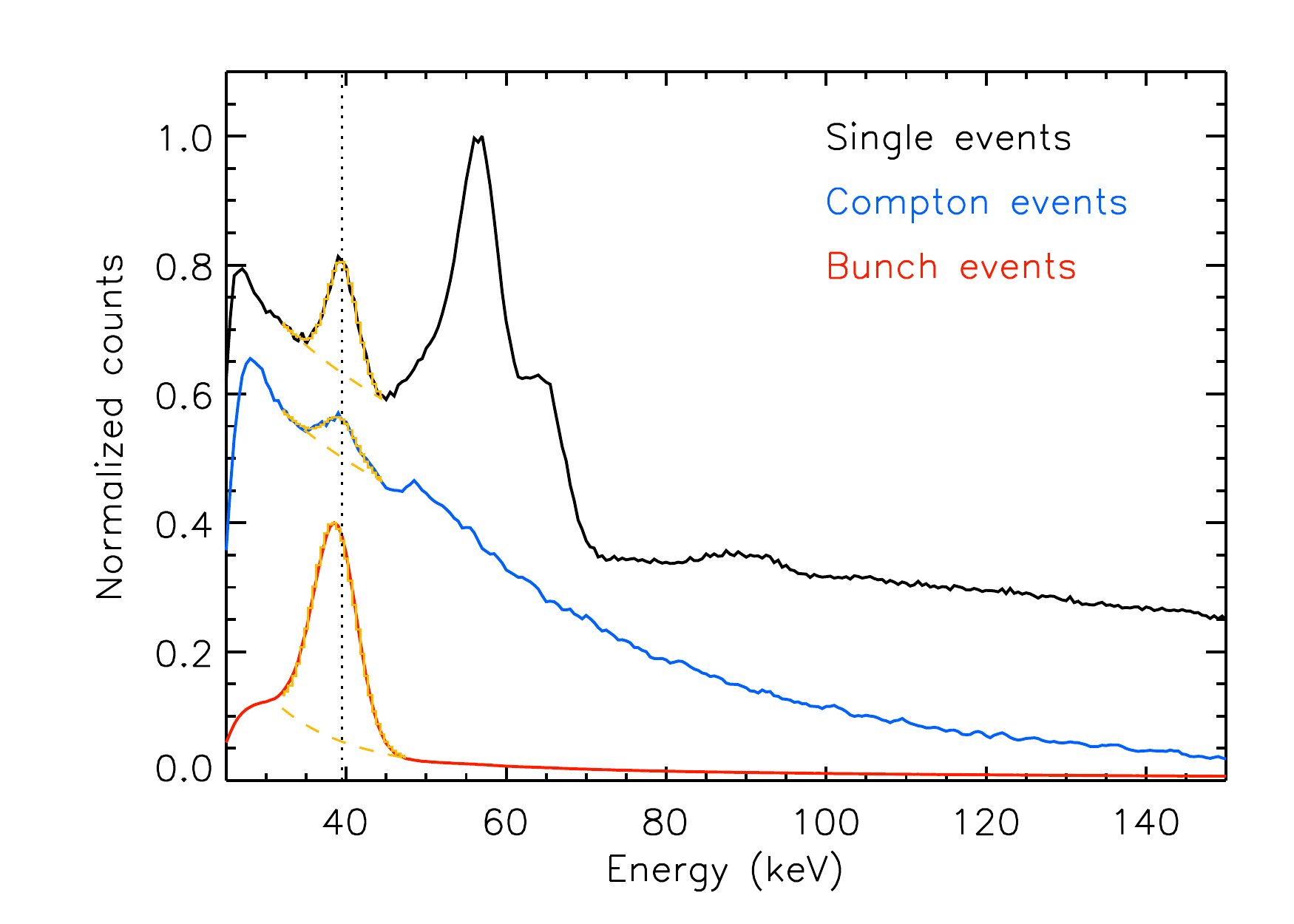}
\caption{The spectra of single and compton double events showing peak around 40 keV as also seen in the `bunches' indicating the particle origin of these events.}
\label{spec_nc}
\end{figure}

\begin{table*}[htb!]
\caption{Signal to noise ratio (S/N) and 3 $\sigma$ outliers per quadrant calculated for 11 GRBs using the \textit{cztpipeline\_lowthresh\_run}, \textit{cztpipeline\_highthresh\_run} and new event selection algorithm. For extremely bright GRB160821A \textit{cztpipeline\_lowthresh\_run} shows a negative S/N since the counts during the GRB interval was killed and instead of a peak it resulted in a drop in the lightcurve.}
\label{snr_grb}
\centering
\renewcommand{\arraystretch}{1.2}
\resizebox{\textwidth}{!}{
\begin{tabular}{c c c c c c c} 
\hline\hline
GRB name &\multicolumn{2}{c}{\textit{cztpipeline\_lowthresh\_run}} & \multicolumn{2}{c}{\textit{cztpipeline\_highthresh\_run}} & \multicolumn{2}{c}{new algorithm} \\
 & S/N & 3$\sigma$ outliers  & S/N & 3$\sigma$ outliers & S/N & 3$\sigma$ outliers \\
\hline\hline
GRB191225B & $2.07$ & $2.76$ & $2.01$ & $5.35$ & $2.3$ & $1.52$\\ 
GRB190315A & $4.24$ & $2.52$ & $5.3$ & $4.01$ & $5.54$ & $0.85$\\ 
GRB180703B & $3.36$ & $1.51$ & $6.64$ & $3.06$ & $6.97$ & $1.08$\\ 
GRB180504B & $3.89$ & $1.69$ & $4.15$ & $3.48$ & $4.22$ & $0.94$\\ 
GRB180120A & $8.76$ & $4.48$ & $11.57$ & $3.45$ & $11.99$ & $0.92$\\ 
GRB180411C & $5.01$ & $3.06$ & $5.09$ & $5.92$ & $5.45$ & $2.08$\\ 
GRB171126A & $4.13$ & $3.76$ & $4.15$ & $2.99$ & $4.34$ & $1.25$\\ 
GRB171223A & $3.9$ & $2.02$ & $5.75$ & $5.49$ & $6.23$ & $0.91$\\ 
GRB200916B & $3.97$ & $2.88$ & $5.54$ & $2.09$ & $6.43$ & $1.19$\\ 
GRB200920B & $3.41$ & $0.98$ & $2.95$ & $3.73$ & $3.25$ & $1.21$\\ 
GRB160821A & $-0.51$ & $1.76$ & $35.67$ & $2.88$ & $38.2$ & $1.41$\\ 
\hline
\hline
\end{tabular}
}
\end{table*}

Now, we examine the spectra of cleaned events. Fig. \ref{spec_nc} shows the energy spectrum for clean single pixel events for ObsID2, along with spectrum of the scattered events of the double pixel Compton events that are used for polarimetry. The single event spectrum shows the expected Tantalum $K-\alpha$ and $K-\beta$ lines and an additional line feature around 40 keV. This line feature at 40 keV is also seen in Compton event spectrum and is known to be a prominent feature in the events that constitute the ‘bunches’, as shown by the bunch event spectrum in the figure. Thus, a small fraction of events having similar spectral characteristics as the bunch events remain present in the clean event files. To estimate the contribution of such events, the spectra around the line were fitted with a powerlaw continuum and a Gaussian line and fraction of events within the line were computed. It is seen that about 1-1.2 \% of the total clean single and Compton events are arising from this line at 40 keV and the line fraction is about 40 \% for bunch events. Assuming that the spurious events left in the clean event files have spectrum similar to the bunch events, we estimate the fraction of residual noise events to be $\sim$ 2.5-3 \%. Although the effect of these residual events are visible in the energy spectrum, they do not show any clustering in time or detector position, which is why they are not removed even after the improved event filtering techniques. As these residual events are random in time and uniform over the detector plane, they act as additional background events and thus get removed in background subtraction. Thus, this small fraction of residual events do not contribute to any additional systematic errors.

\subsection{Signal to noise of GRBs}
%\begin{figure*}[ht!]
%\centering
%\begin{subfigure}[b]{0.4\textwidth}
%\centering
%\includegraphics[width=\textwidth]{GRB180411C.png}
%\caption{GRB171126A in quad 2}
%\label{}
%\end{subfigure}
%\hfill
%\begin{subfigure}[b]{0.4\textwidth}
%\centering
%\includegraphics[width=\textwidth]{GRB180504B.png}
%\caption{GRB180703B in quad 2}
%\label{}
%\end{subfigure}
%\hfill
%\begin{subfigure}[b]{0.4\textwidth}
%\centering
%\includegraphics[width=\textwidth]{GRB180703B.png}
%\caption{GRB191225B in quad 3}
%\label{}
%\end{subfigure}
%\caption{Lightcurves of different GRBs from the new event selection algorithm (green), and %\textit{cztpipeline\_highthresh\_run} (blue). The vertical dotted line indicates the GRB trigger time. The outliers seen in the blue and not in the green are the peaks due to particle induced noises.}
%\label{GRB_lightcurves}
%\end{figure*}

\begin{figure*}[p!]
\centering
\subfloat[GRB180504B Quadrant 0]{
\label{subfig:correct}
\includegraphics[width=0.8\textwidth]{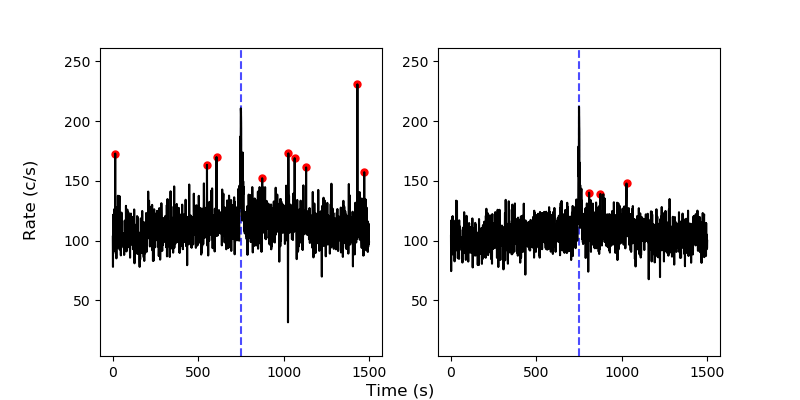}}
	
\subfloat[GRB180703B Quadrant 1]{
\label{subfig:correct}
\includegraphics[width=0.8\textwidth]{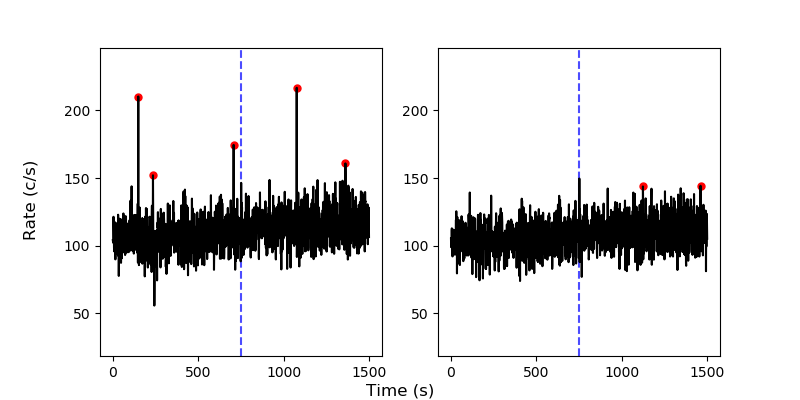}}
	
\subfloat[GRB191225B Quadrant 0]{
\label{subfig:correct}
\includegraphics[width=0.8\textwidth]{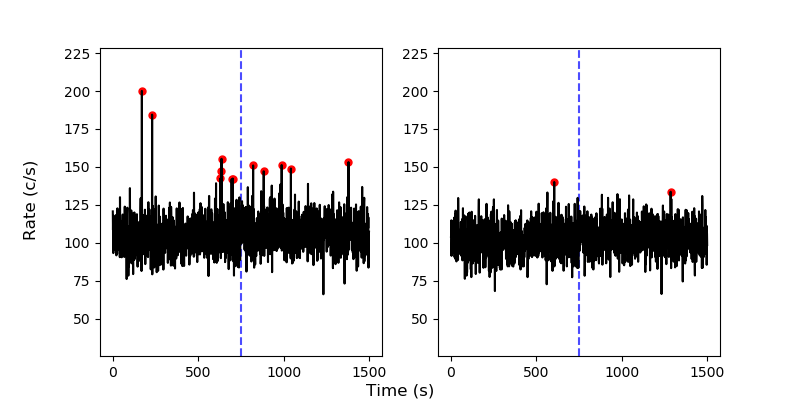}}
\caption{CZTI Lightcurves of different GRBs at quadrant level from \textit{cztpipeline\_highthresh\_run} (left) and the new event selection algorithm (right). The vertical blue dotted line indicates the GRB trigger time. The 3 $\sigma$ outliers are marked by red dots and are seen more in the \textit{cztpipeline\_highthresh\_run}.}
\label{GRB_lightcurves}
\end{figure*}

We calculated the signal to noise ratio in 11 gamma ray bursts (GRBs), previously detected by CZTI to show the reduction in the background noise without compromising on the source counts for the new algorithm. We compare our results with the \textit{cztpipeline\_lowthresh\_run} and \textit{cztpipeline\_highthresh\_run} of the cztpipeline. The light curve of the short and long GRBs were binned at 1 s. Bins with exposure times less than 0.3 are excluded from the analysis. The trigger time (T$_{trig}$) are obtained from the AstroSat CZTI GRB catalog (\href{http://astrosat.iucaa.in/czti/?q=grb}{http://astrosat.iucaa.in/czti/?q=grb}). The source region in the light curve was selected manually. 500 seconds before and after the GRB region was selected as background. \\
%Similarly for long GRB 100 seconds before and after GRBs was selected as background.

However the background pre-and-post GRB background was taken 5 seconds away from the GRB to avoid contamination from the source counts in modelling the background. Since the background of CZTI follows a trend we fit a quadratic function to de-trend the pre-and-post GRB backgrounds for all the quadrants. The fitted function was then subtracted from the light curve to obtain a background subtracted light curve. The signal to noise ratio of the GRB is estimated by M/S, where M is the mean during the background subtracted GRB time interval, and S is the standard deviation of the de-trended background. The sample consists of 5 short GRBs (T$_{90}$ $<$ 2 s) and 6 long GRBs (T$_{90}$ $>$ 2 s). \\

%\begin{table}[hb!]
%\caption{Signal to noise ratio (SNR) calculated for 11 GRBs using the \textit{cztpipeline\_highthresh\_run} run of the pipeline and new event selection algorithm.}
%\label{snr_grb}
%\centering   
%\renewcommand{\arraystretch}{1.5}
%\begin{tabular}{c c c} 
%\hline\hline  
%GRB name& SNR & SNR \\
%    & \textit{cztpipeline\_highthresh\_run} & new algorithm %\\
%\hline 
%GRB191225B &	1.87  &	2.15 \\
%GRB190315A &	5.73  &	5.90 \\
%GRB180703B &	6.88  &	7.14 \\
%GRB180504B &	3.71  &	4.00 \\
%GRB180120A &	11.36 &	12.16 \\
%GRB180411C &	4.47  &	4.94 \\
%GRB171126A & 	3.01  &	3.24 \\
%GRB171223A &	9.08  &	9.44 \\
%GRB200916B &	3.78  &	4.05 \\
%GRB200920B &	2.03  &	2.05 \\
%\hline
%\hline
%\end{tabular}
%\end{table}
Table.\ref{snr_grb} shows the signal to noise ratio of all 11 known GRBs in comparison with the \textit{cztpipeline\_highthresh\_run} and \textit{cztpipeline\_lowthresh\_run}. We also calculated the amount of 3 $\sigma$ peaks per quadrant above the poisson limit. This is calculated by subtracting the estimated amount of peaks beyond 3 $\sigma$ level from the Poissonian equation from the observed peaks in the lightcurve beyond the 3 $\sigma$ level at the quadrant level. Table. \ref{snr_grb} shows that while the signal have been lost in some cases for the \textit{cztpipeline\_lowthresh\_run}, 3 $\sigma$ peaks per quadrant increase in case of the \textit{cztpipeline\_highthresh\_run}. However the new algorithm retains the GRB signal without compromising on the amount of particle induced 3 $\sigma$ peaks. It is also seen that there is a slight improvement of the signal to noise ratio, as well as a reduction in the particle peaks in comparison to both the methods from cztpipeline. Fig. \ref{GRB_lightcurves} shows the lightcurves of the GRB from the \textit{cztpipeline\_highthresh\_run} of cztpipeline and after noise clean algorithm. The peaks that disappear after the new event selection algorithms are the particle induced noises. It is evident from the lightcurves, that outliers in lightcurves in second and sub-second time scales has reduced significantly, and the search for astrophysical transients will not be influenced by particle induced and electronic pixel noises. Hence the false alarm rates due to particle induce noises will be drastically reduced for an automatic transient search.\\

\section{Conclusions}
In this paper we have outlined a method of event selection in the CZTI data based on their clustering properties at spatial, spectral and temporal dimensions. It is found that 
the cosmic ray induced  noise in the CZTI data has been reduced by a significant amount without source flux dependent manual configuration of the parameters. The same algorithm provides cleaned data for both highly variable short duration bright transient sources as well as for the background dominated  on-axis source observations,  with the same configuration of the parameters. We also found that the cosmic ray tracks which can mimick short GRBs are also reduced significantly. Searches for transients, especially short transients, like short GRBs, counterparts to gravitational wave events, counterparts to fast radio bursts (FRBs), will benefit by using this algorithm as the false trigger rate will be reduced significantly. 
%The significant improvement in the pulsed counts for an off axis Crab pulsar observation with respect to \textit{lowthreshconfiguration} means that fraction of genuine X-ray has increased. This shows the significant improvement in sensitivity for off axis sources. 
The reduction in the particle tracks not only improves the transient and GRB searches but also the science products like spectrum, polarisation and localisation for them. This is because the particle tracks are not uniformly distributed across time for short timescales like a few hundreds of seconds, which makes the background subtraction in energy and DPH improper. Since the new algorithm is robust at divergent count levels, it will be very useful for combining data obtained for long duration at different flux levels like that needed for  the off axis pulsar searches, especially for the fainter milli-second pulsars. Further co-adding data to search for sources at lower flux levels is easier with this new algorithm because the parameters can be set uniformly across data sets and an automatic analysis procedure can be established. \\

%Since the amount of noise reduced is constant in comparison to sources of different flux levels, the improvement in the capability to detect  low flux sources will be very significant. This will greatly improve the off axis pulsar searches, especially for the fainter milli-second pulsars.
With additional features in flagging `noisy' and `flickering' pixels, this algorithm has a better handling of the pixelated noises in temporal and spatial regimes in comparison to the cztipipeline.  Currently a beta version of the algorithm is available. This will be used in multiple data sets for different uses. The algorithm is organised in a structured way such that the default parameters can be tweaked, the order and sequence of analysis can be experimented such that a robust understanding of the various aspects of noise could be arrived at. A new version of cztipipeline is currently being developed to include improved aspects of energy calibration and imaging and we plan to incorporate this noise clean algorithm in the new pipeline.
%In future we will also explore the improvement in signal to noise ratio of the polarisation measurements. We also plan to include this algorithm in the upcoming versions of the cztpipeline. Apart  from improving the results of CZTI, this algorithm provides a premise for noise reduction in future hard X-ray open sky detectors using CdZnTe crystals.  

%%Use section* for acknowledgements
\section*{Acknowledgements}
The data used in this work is from the {\em AstroSat} mission of the Indian Space Research Organization (ISRO), archived at the Indian Space Science Data Centre (ISSDC). We acknowledge the {\em AstroSat} CZTI Payload operation centre at IUCAA for their help in developing and testing this algorithm. 
The CZT Imager instrument was built by a TIFR-led consortium of institutes across India, including VSSC, ISAC, IUCAA, SAC, and PRL. The Indian Space Research Organisation funded, managed and facilitated the project.
We thank Mayuri Shinde for her contributions in developing the algorithm into a code in C++ programming language. We also thank Avinash Aher for his contributions in data analysis.

\vspace{-1em}

%%use \balance somewhere in the left column of the last page to balance the two columns in the end page

%%References section

\bibliographystyle{apj}
\bibliography{references_noiseclean}

\end{document}